\documentclass[12pt]{utarticle-dan}

\pdfoutput=1

\usepackage{amsmath,amsthm,amsfonts,amscd,amssymb} 
\usepackage{eucal}
\usepackage{braket}
\usepackage{hyperref}
\usepackage{graphicx}

\newcommand{\Hi}{\mathcal{H}}
\newcommand{\Oi}{\mathcal{O}}

\newcommand{\be}{\begin{equation}}
\newcommand{\ee}{\end{equation}} 
\newcommand{\mb}{\mathbf}

\newcommand{\Li}{\mathcal{L}}

\newcommand{\ul}{\underline}
\newcommand{\bear}{\begin{eqnarray}}
\newcommand{\eear}{\end{eqnarray}}

\DeclareMathOperator{\tr}{tr}

\begin{document}

\title{Scattering with partial information}

    \author{Daniel Carney, Laurent Chaurette, and Gordon Semenoff
     \oneaddress{
	  Department of Physics and Astronomy\\
	  University of British Columbia\\
	  Vancouver, BC V6T 1Z1 Canada \\
	  {~}\\
	  \email{carney@phas.ubc.ca}\\
	  \email{dodeca@phas.ubc.ca}\\
	  \email{gordonws@phas.ubc.ca}\\
       }
   }

\Abstract{We study relativistic scattering when one only has access to a subset of the particles, using the language of quantum measurement theory. We give an exact, non-perturbative formula for the von Neumann entanglement entropy of an apparatus particle scattered off an arbitrary set of system particles, in either the elastic or inelastic regime, and show how to evaluate it perturbatively. We give general formulas for the late-time expectation values of apparatus observables. Some simple example applications are included: in particular, a protocol to verify preparation of coherent superpositions of spatially localized system states using position-space information in the outgoing apparatus state, at lowest order in perturbation theory in a weak apparatus-system coupling.}

\maketitle

\newpage

\tableofcontents

\section{Introduction}

The purpose of this paper is to make contact between concepts from quantum information and relativistic scattering theory. In particular, we study how to use interacting fields as measurement devices.

In standard formulations of measurement theory, one imagines performing a measurement of a system $S$ by coupling it to an apparatus $A$. We start the apparatus in some register state $\ket{0}_A$ while the system is in an arbitrary superposition, and then entangle these in such a way that measurements on $A$ can determine the initial state of $S$. Schematically, one writes things like
\be
\ket{0}_A \otimes \sum_i c_i \ket{i}_S \to  \sum_i c_i \ket{i}_A \otimes \ket{i}_S,
\ee
with the arrow referring to time evolution under some total Hamiltonian (see eg. \cite{JvNbook,preskillnotes}). This process necessarily generates  entanglement between $S$ and $A$. The goals of this paper are to study to what extent we can understand the scattering of system particles $S$ by another particle $A$ in this language and to quantify how much entanglement is generated in such scattering events.

To this purpose, we consider an arbitrary system of fields and append an apparatus field $\phi_A$ which we can scatter off the system, so we consider Hilbert spaces formed by tensor products of apparatus and system fields. The $S$-matrix generates entanglement between the factors. This approach differs from and complements other ways of dividing field-theoretic systems; one can also consider, for example, divisions by spatial area \cite{Calabrese:2004eu,Ryu:2006bv}, momentum scale \cite{Balasubramanian:2011wt}, or multiple non-interacting CFTs \cite{Maldacena:2001kr}.

We begin by reviewing and slightly extending the textbook treatment \cite{Weinberg:1995mt} of scattering theory to incorporate density matrices as initial conditions in section \ref{generalformalismsection}. We explain how to calculate expectation values of operators probing only the apparatus. In section \ref{entropysection}, we present an exact, non-perturbative formula for the von Neumann entropy of the apparatus $A$ after the scattering event, assuming only that the state at early and late times contains exactly one particle of $\phi_A$.

We then apply these results to the simplest possible example, in which the apparatus and system both consist of a single particle of some scalar fields $\phi_{A,S}$, with $A$ and $S$ weakly coupled. In section \ref{simplesection} we give an explicit formula for the entropy generated when we scatter a product momentum state $\ket{\mb{p}}_A \ket{\mb{q}}_S$, recovering and slightly correcting a result of \cite{Park:2014hya,Peschanski:2016hgk}.

In section \ref{superpositionsection}, we consider a somewhat different problem. Suppose that we think we are preparing the system $S$ in a superposition of two well-localized position states. We show how to do a measurement with $A$ to verify that the superposition is really coherent, as opposed to (say) having decohered into a classical ensemble. We find that a good observable to use to determine the coherence of $S$ is position-space interference fringes in the outgoing distribution for the apparatus particle $A$. These show up at lowest order in perturbation theory in the $S$-$A$ coupling $\lambda$, whereas the momentum-space distribution of $A$ is only sensitive at second order.

\section{Scattering with density matrices}

\label{generalformalismsection}

\subsection{General considerations}

Let's consider the general problem of scattering where we know the state of the total system at very early times $t \to -\infty$, and we want to know how this evolves at very late times due to a scattering event. We want to consider any density matrix for the full system as an initial condition. The treatment here is a straightforward generalization of Weinberg's textbook \cite{Weinberg:1995mt}, and our conventions throughout follow his. In particular, the metric signature is $-+++$ and $\hbar = c = 1$. 

Assume the total Hamiltonian can be written
\be
H = H_0 + V,
\ee
and denote the energy eigenstates of the free Hamiltonian $H_0$ as
\be
\label{freeeigenvals}
H_0 \ket{\alpha} = E_{\alpha} \ket{\alpha}.
\ee
Here the label $\alpha = \mb{p}_1 \sigma_1 n_1, \mb{p}_2 \sigma_2 n_2, \ldots$ covers the momentum, spin, and particle species of the free-particle states. We define in- and out-states as Heisenberg-picture states which have the energies $E_{\alpha}$ but are eigenstates of the \emph{full} Hamiltonian,\footnote{Notice that the conditions \eqref{freeeigenvals} and \eqref{inoutdef1} mean that the ``free'' states and scattering states have the same energy spectrum. This means in particular that the masses appearing in the Hamiltonian are the physical (``renormalized'' or ``dressed'') masses of the particles.}
\be
\label{inoutdef1}
H \ket{\alpha^{\pm}} = E_{\alpha} \ket{\alpha^{\pm}},
\ee
satisfying the condition that as $t \to \mp \infty$, for any reasonably smooth functions $g^{\pm}(\alpha)$ of the particle labels,
\be
\label{inoutdef2}
\ket{\psi} = \int d\alpha \ e^{-i E_{\alpha} t} g^{\pm}(\alpha) \ket{\alpha^{\pm}} \to \int d\alpha \ e^{-i E_{\alpha} t} g^{\pm}(\alpha) \ket{\alpha}.
\ee 
This condition says that at very early or late times, the in/out states behave like the free-particle states of the corresponding particle labels $\alpha$. The notation is that $+$ indicates an in-state while $-$ denotes an out-state. Both the free and scattering states are taken to be Dirac delta-normalizable $\braket{\alpha | \alpha'} = \braket{\alpha^{\pm} | \alpha^{'\pm}} = \delta(\alpha-\alpha')$.

If the system is in a wavepacket like \eqref{inoutdef2}, and we know the matrix elements $\braket{\alpha | \Oi | \alpha'}$ of some observable in terms of free-particle states, we can compute the expectation value of $\Oi$ at early or late times in the state $\ket{\psi}$ as follows. In the Heisenberg picture we have $\Oi(t) = e^{i H t} \Oi e^{-i H t}$, so using \eqref{inoutdef1} and \eqref{inoutdef2}, we have that as $t \to \mp \infty$,
\be
\braket{ \psi | \Oi(t) | \psi} \to \int d\alpha d\alpha' \ e^{i(E_{\alpha} - E_{\alpha'})t} g^*(\alpha) g(\alpha') \braket{ \alpha | \Oi | \alpha'}.
\ee
More generally, the system may be in a density matrix. This can be decomposed into any complete basis, including the scattering states:
\be
\rho = \int d\alpha d\alpha' \rho^{\pm}(\alpha,\alpha') \ket{\alpha^{\pm}} \bra{\alpha^{\pm}}.
\ee
Then the expectation value of $\Oi$ is given asymptotically by
\be
\label{obsout}
\braket{ \Oi(t) } = \tr \rho \Oi(t) \to \int d\alpha d\alpha' \ \rho^{\pm}(\alpha,\alpha') e^{i(E_{\alpha} - E_{\alpha'})t} \braket{\alpha | \Oi | \alpha'}
\ee
as $t \to \mp \infty$.

Since the states $\ket{\alpha^+}$ and $\ket{\alpha^-}$ separately form complete bases for positive-energy states of the system, we can express one base in terms of the other. The $S$-matrix is the unitary operator with elements given by the inner product
\be
S_{\beta \alpha} = \braket{ \beta^- | \alpha^+}.
\ee
The in- and out-coefficients of the density matrix are thus related by
\be
\label{DMout}
\rho^-(\beta,\beta') = \int d\alpha d\alpha' \ S_{\beta \alpha} S^*_{\beta' \alpha'} \rho^+(\alpha,\alpha').
\ee
We will always consider Poincar\'{e}-invariant systems. We can therefore write the $S$-matrix as an identity term plus a term with the total four-momentum invariance factored out,
\be
\label{Sdecomp}
S_{\beta\alpha} = \delta(\beta - \alpha) - 2\pi i M_{\beta\alpha} \delta^4(p_{\beta} - p_{\alpha}).
\ee
In appendix \ref{opticaltheoremapp}, we use the unitarity of the $S$-matrix,
\be
\label{Smatrixunitarity}
\int d\beta S_{\beta \alpha} S^*_{\beta \alpha} = \delta(\alpha-\alpha')
\ee
to derive the optical theorem, \eqref{opticaltheorem}, which will play a role repeatedly in the calculations that follow.

\subsection*{Box normalizations}
In computing various quantities it will be useful to work with discrete states. We can do this by putting the entire process into a large spacetime volume of duration $T$ and spatial volume $V = L^3$. Periodic boundary conditions on $V$ allow us to retain exact translation invariance. We define dimensionless, box-normalized states
\be
\label{boxregs}
\ket{\alpha^{\pm}}^{box} = \tilde{N}^{n_\alpha/2} \ket{\alpha^{\pm}}, \ \ \tilde{N} =\frac{(2\pi)^3}{V},
\ee
where $n_{\alpha}$ is the number of particles in the state $\alpha$. When working directly with box-normed states, delta functions and $S$-matrix elements are all dimensionless, integrals over states are replaced by sums, and the delta-functions are Kroneckers. We have
\be
S^{box}_{\beta\alpha} = \tilde{N}^{(n_{\alpha}+n_{\beta})/2} S_{\beta\alpha},
\ee
by definition of the $S$-matrix. Delta functions are then regulated as
\be
\label{deltaregs}
\delta_V^3(\mb{p}-\mb{p}') = \tilde{N}^{-1} \delta_{\mb{p},\mb{p}'}, \ \ \ \delta_T(E-E') = \frac{1}{2\pi} \int_{-T/2}^{T/2} dt \ e^{i (E - E') t}.
\ee
Note in particular that this implies $\delta_T(0) = T/2\pi$. We then define a box-normalized transition amplitude:
\be
\label{Sdecompbox}
S^{box}_{\beta\alpha} = \delta_{\beta \alpha} - 2\pi i M^{box}_{\beta \alpha} \delta_{\mb{p}_{\beta} \mb{p}_{\alpha}} \delta_{E_{\beta} E_{\alpha}} \iff M^{box}_{\beta\alpha} = \tilde{N}^{(n_{\alpha}+n_{\beta}-2)/2} M_{\beta\alpha}.
\ee
Note that $M^{box}$ has mass dimension one, since $\delta_T(E)$ has dimensions of inverse mass.

\subsection{Measuring the apparatus state}

Suppose now that we divide the total system into an apparatus $A$ and system $S$ and only have direct access to $A$. Here we work out a formula for computing observables only of $A$, and for the von Neumann entropy of $A$.

In what follows, we assume that $A$ and $S$ are distinguishable; a simple way to achieve this is to just have $A$ and $S$ described by different fields. We will make this assumption in everything that follows. We will hereafter make a slight abuse of the previous notation and label states with two indices $(a, \alpha)$ where $a$ labels apparatus eigenstates and $\alpha$ labels system eigenstates. We can decompose the total Hilbert space as a product over free, in, or out states:
\be
\label{hilbertdecomp}
\Hi = \Hi_A \otimes \Hi_S = \Hi_A^{\pm} \otimes \Hi_S^{\pm}.
\ee
The total $S$-matrix provides a unitary map between the in- and out-state decompositions. In particular, a product in-state is a generally non-separable mixture of out-states:
\be
\ket{a \alpha}^+ = \int db d\beta \ S_{b\beta,a\alpha} \ket{b\beta}^-.
\ee

At early or late times, we want to compute the expectation value of any observable $\Oi_A : \Hi_A \to \Hi_A$. Note that here $\Oi_A$ is an operator on the \emph{free} apparatus Hilbert space factor in \eqref{hilbertdecomp}. Take $\Oi = \Oi_A \otimes \mb{1}_S$ and apply \eqref{obsout}. By the asymptotic conditions on the scattering states, a simple calculation shows that at early or late times
\be
\label{redobs3}
\braket{ \Oi_A(t) } := \braket{\Oi(t)} \to \int da da' d\alpha \ \rho^{\pm}(a,\alpha;a',\alpha) e^{i(E_{a} - E_{a'})t} \braket{a | \Oi_A | a'}.
\ee
To derive this formula, we assumed that the free Hamiltonian has an additive spectrum $H_0 \ket{a \alpha} = (E_a + E_{\alpha}) \ket{a \alpha}$. The result \eqref{redobs3} holds for any density matrices; in particular, we do not need to assume that the total state factors into a product of a density matrix for $A$ and a density matrix for $S$ at either early or late times.

We would also like to define the entanglement entropy between apparatus and system. To do this, we again use the decomposition \eqref{hilbertdecomp} to perform partial traces over the system. We can do this using either in- or out-states,
\be
\rho^{\pm}_A := \tr_{\Hi_S^{\pm}} \rho
\ee
from which we can in turn define the entanglement entropy
\be
\label{EEdef}
S_A^{\pm} = - \tr_{\Hi_A^{\pm}} \rho_A^{\pm} \ln \rho_A^{\pm}.
\ee

\section{$A$-$S$ entanglement entropy}
\label{entropysection}

Our goal in this section is to calculate the entanglement entropy between the system and apparatus at late times. Consider the system and apparatus both prepared in definite momentum eigenstates at early times,
\be
\ket{\psi} = \ket{\mb{p}^+}_A \ket{\alpha^+}_S.
\ee
Here as before $\alpha = \mb{q}_1 n_1 \sigma_1, \mb{q}_2 n_2 \sigma_2, \ldots$ labels all the momenta, species, and spin of the system particles, while $\mb{p}$ is simply the initial momentum of the apparatus, which we take to be a scalar for notational simplicity. For the entirety of this section until the end, we will work in a spacetime box as described above, but will refrain from writing ``box'' superscripts. At the end of the computation we will discuss the continuum limit.

We assume that one and only one apparatus particle exists in both the initial and final state. This can be arranged for example by assigning $\phi_A$ some global charge, or by taking $\phi_A$ to have high mass and studying scattering events below its production threshold. 

\begin{figure}
\centering
\includegraphics{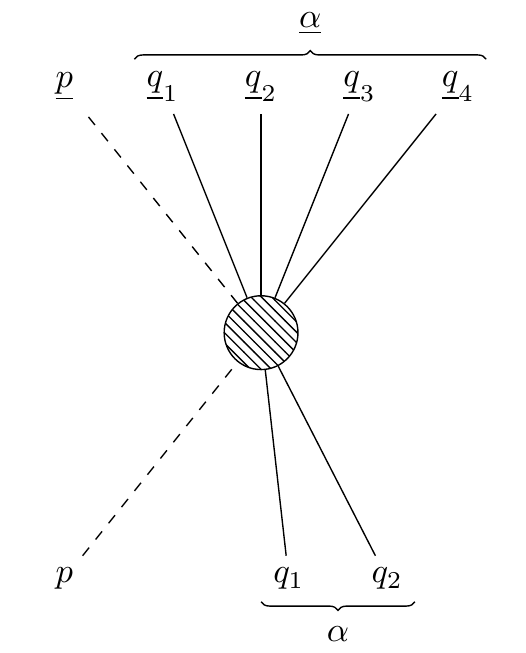}
\caption{A typical apparatus-system scattering process. Dotted lines denote the apparatus, solid lines the system. Time runs from bottom to top.}
\label{scatterpic}
\end{figure}

Using the formalism from section \ref{generalformalismsection}, we can express the density matrix in terms of out-states,
\be
\rho = \sum_{\ul{\mb{p}} \ul{\mb{p}}' \ul{\alpha} \ul{\alpha}'} S_{\ul{\mb{p}} \ul{\alpha} \mb{p} \alpha}  S^*_{\ul{\mb{p}}' \ul{\alpha}' \mb{p} \alpha} \ket{\ul{\mb{p}} \ul{\alpha}^-} \bra{\ul{\mb{p}}' \ul{\alpha}^{'-}}.
\ee
From here out we use underlines to denote outgoing variables. Expanding the $S$-matrix with \eqref{Sdecompbox}, one can see from this expression that $\rho$ will have the correct norm $\tr \rho = 1$ if and only if the optical theorem \eqref{opticaltheorem} is satisfied (see appendix \ref{opticaltheoremapp}). In particular, if one is working in perturbation theory, the optical theorem mixes orders, so one needs to be careful about including the correct set of loop and tree diagrams at a given order.

Now trace over the system, using out-states:
\be
\rho_A^- = \sum_{\ul{\mb{p}} \ul{\mb{p}}' \ul{\alpha}} S_{\ul{\mb{p}} \ul{\alpha} \mb{p} \alpha}  S^*_{\ul{\mb{p}}' \ul{\alpha} \mb{p} \alpha} \ket{\ul{\mb{p}}^-} \bra{\ul{\mb{p}}^{'-}}.
\ee
Decompose the $S$-matrix with \eqref{Sdecompbox}. We get three types of terms: from the delta-squared we get a term on the diagonal with momentum given by the initial momentum $\mb{p}$:
\be
\rho_{A,1}^- = \ket{\mb{p}^-} \bra{\mb{p}^-}.
\ee
The cross-terms $-i M \rho + i \rho M^{\dagger}$ give a contribution
\begin{align}
\begin{split}
\rho_{A,2}^- = -2 T \ \text{Im}\left[ M_{\mb{p}\alpha\mb{p}\alpha} \right] \ket{\mb{p}^-} \bra{\mb{p}^-},
\end{split}
\end{align}
again to the density matrix element for the initial momentum $\mb{p}$. This is the forward scattering term that appears in the optical theorem. Finally, we need the terms from $M \rho M^{\dagger}$. One obtains
\begin{align}
\begin{split}
\rho_{A,3}^- = 2\pi T \sum_{\ul{\mb{p}} \ul{\alpha}} \left| M_{\ul{\mb{p}} \ul{\alpha} \mb{p} \alpha} \right|^2 \delta_{\ul{\mb{p}} + \mb{p}_{\ul{\alpha}}, \mb{p} + \mb{p}_{\alpha}} \delta(E^A_{\ul{\mb{p}}}+E^S_{\ul{\alpha}} - E^A_{\mb{p}} - E^S_{\alpha}) \ket{\ul{\mb{p}}^-} \bra{\ul{\mb{p}}^{-}}.
\end{split}
\end{align}

We see that the reduced density matrix for $A$ is diagonal in an arbitrary reference frame. This is due entirely to translation invariance and our assumption that we always have precisely one apparatus particle. Writing the apparatus state in matrix form, we have
\be
\rho_A^- = \begin{pmatrix} 1 + I_0 + F(\mb{p}) & & & \\
& F(\ul{\mb{p}}_1) & &  \\
& &  F(\ul{\mb{p}}_2) & \\
& & & \ddots \end{pmatrix},
\ee
where the $\ul{\mb{p}}_i$ are all the outgoing apparatus momenta $\ul{\mb{p}} \neq \mb{p}$. The coefficients are
\begin{align}
\begin{split}
\label{entropyboxcoeffs}
I_0 & = - 2T \ \text{Im} M_{\mb{p}\alpha \mb{p}\alpha} \\
F(\ul{\mb{p}}) & = 2\pi T \sum_{\ul{\alpha}} \left| M_{\ul{\mb{p}} \ul{\alpha} \mb{p} \alpha} \right|^2 \delta_{\ul{\mb{p}} + \mb{p}_{\ul{\alpha}}, \mb{p} + \mb{p}_{\alpha}} \delta(E^A_{\ul{\mb{p}}}+E^S_{\ul{\alpha}} - E^A_{\mb{p}} - E^S_{\alpha}).
\end{split}
\end{align}
The coefficients $F(\ul{\mb{p}})$ could be called ``conditional transition probabilities''. They are given by fixing an apparatus out-momentum $\ul{\mb{p}}$ and then summing over the transition probabilities to all the possible system states consisent with total momentum conservation. Note that $F(\ul{\mb{p}}) = 0$ for momenta violating energy conservation, that is when $E^A_{\ul{\mb{p}}} > E^A_{\mb{p}} + E^S_{\alpha} - E^S_0$.\footnote{In $2 \to 2$ scattering, we can write the return-amplitude term $I_0 + F(\mb{p})$ in a way that treats the two particles more symmetrically: by the optical theorem \eqref{opticaltheorem}, we have
\be
I_0 = -(2\pi)^2 \sum_{\ul{\mb{p}} \ul{\mb{q}}} \left| M_{\ul{\mb{p}} \ul{\mb{q}} \mb{p} \mb{q}} \right|^2 \delta_{\ul{\mb{p}} + \ul{\mb{q}}, \mb{p} + \mb{q}} \delta_{E^A_{\ul{\mb{p}}}+E^S_{\ul{\mb{q}}},E^A_{\mb{p}}+E^S_{\mb{q}}}
\ee
while by definition, $F(\mb{p}) = (2\pi)^2 \left| M_{\mb{p} \mb{q} \mb{p} \mb{q}} \right|^2$. So the shift in the initial-momentum density matrix eigenvalue is
\be
\Delta_0 = -(I_0 + F(\mb{p})) = (2\pi)^2 \sum_{(\ul{\mb{p}},\ul{\mb{q}}) \neq (\mb{p},\mb{q})} \left| M_{\ul{\mb{p}} \ul{\mb{q}} \mb{p} \mb{q}} \right|^2 \delta_{\ul{\mb{p}} + \ul{\mb{q}}, \mb{p} + \mb{q}} \delta_{E^A_{\ul{\mb{p}}}+E^S_{\ul{\mb{q}}},E^A_{\mb{p}}+E^S_{\mb{q}}}.
\ee}

The von Neumann entanglement entropy of the apparatus is given by
\be
\label{entropyboxformula}
S_A = -(1+I_0 + F(\mb{p})) \ln (1 + I_0 + F(\mb{p})) - \sum_{\ul{\mb{p}} \neq \mb{p}} F(\ul{\mb{p}}) \ln F(\ul{\mb{p}}).
\ee
The result \eqref{entropyboxformula} is exact and non-perturbative. It follows completely from Lorentz invariance and our assumption that precisely one $A$ particle is in both the initial and final state. It can be simplified by invoking perturbation theory: we assume that the scattering amplitudes are significantly less than unity. Then $|I_0 + F(\mb{p})| \ll 1$, so we can Taylor expand the first term in \eqref{entropyboxformula} and get a term linear in this expression. But the other terms still have logarithms, so we have an expression like $\text{small} + \sum \text{small} \ln (\text{small})$, and the log terms will dominate. So we are left with
\be
\label{entropyboxformula-PT}
S_A = - \sum_{\ul{\mb{p}}} F(\ul{\mb{p}}) \ln F(\ul{\mb{p}}).
\ee
In a large box, it is immaterial if the sum on outgoing apparatus momenta $\ul{\mb{p}}$ includes $\ul{\mb{p}}=\mb{p}$ or not, since this term is individually of measure zero.

\section{Examples with two scalar fields}

\label{examplesection}

We will now consider some simple applications of the above theory, with both system and apparatus described by scalar fields $\phi_{A,S}$ with a weak coupling $\lambda$. Throughout, we will assume that the initial energies are below the threshold for on-shell pair-production, so that we can work entirely with $2 \to 2$ matrix elements.

In the first subsection, we study entropy generated during a $2 \to 2$ scattering event. In the second subsection, we show how to verify that the system $S$ has been prepared in a spatial superposition by scattering with $A$. More precisely, we show how to read out the coherence of such a superposition using position-space information in $A$, at lowest order in $\lambda$.

Let us fix our conventions. We take the apparatus and system to be described by the action
\begin{align}
\begin{split}
S & = -\int d^4x \ \frac{1}{2} (\partial_{\mu} \phi_S)^2 + \frac{1}{2} (\partial_{\mu} \phi_A)^2 + \frac{1}{2} m_S^2 \phi_S^2 + \frac{1}{2} m_A^2 \phi_A^2 \\
& + \frac{\lambda}{4} \phi_S^2 \phi_A^2  + \frac{\lambda_A}{4!} \phi_A^4 + \frac{\lambda_S}{4!} \phi_S^4 + \Li_{ct}.
\end{split}
\end{align}
In particular, the fields $\phi_{S,A}$ are considered to be distinguishable and renormalized. The term $\Li_{ct}$ contains the counterterms; here we use the standard on-shell renormalization conditions that the on-shell propagators have unit residue at the physical masses and the interactions are given exactly by their physical couplings at threshold. This way we can work with amputated diagrams only, and the lowest order in perturbation theory is just tree level. We will take up loop corrections in a future publication. We assume that the self-couplings $\lambda_{A,S} \ll 1$ and ignore them hereafter.

The free single-particle states and operators are normalized as
\be
\braket{ \mb{k}'  | \mb{k} } = \left[ a_{\mb{k}}, a_{\mb{k}'}^{\dagger} \right] = \delta^{3}(\mb{k}-\mb{k}').
\ee
More generally, a free $n$-particle state of a given species is $\ket{\mb{k}_1 \cdots \mb{k}_n} = a_{\mb{k}_n}^{\dagger} \cdots a_{\mb{k}_1}^{\dagger} \ket{0}$, where $\ket{0}$ is the free vacuum. In what follows we use $\mb{p}$ to denote the 3-momentum of the apparatus and $\mb{q}$ that of the system. The relevant $S$-matrix elements are then
\be
S_{\ul{\mb{p}} \ul{\mb{q}} \mb{p} \mb{q}} = \delta^3(\ul{\mb{p}} -\mb{p}) \delta^3(\ul{\mb{q}} - \mb{q}) - 2 \pi i M_{\ul{\mb{p}} \ul{\mb{q}} \mb{p}\mb{q}} \delta^4(\ul{p} + \ul{q} - p - q)
\ee
with the amplitude given by, to lowest order in perturbation theory,
\begin{align}
\label{22elements}
\begin{split}
i M_{\ul{\mb{p}} \ul{\mb{q}} \mb{p} \mb{q}} & = \vcenter{\hbox{\includegraphics{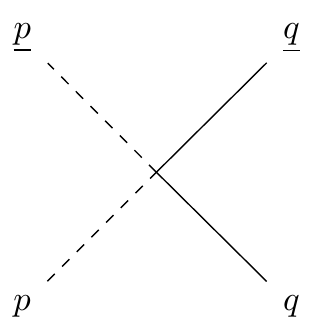}}} = \frac{i \lambda}{(2\pi)^3 \sqrt{16 E^A_{\ul{\mb{p}}} E^S_{\ul{\mb{q}}} E^A_{\mb{p}}  E^S_{\mb{q}}}}.
\end{split}
\end{align}
Here the single-particle energies are
\be
E^{S,A}_{\mb{k}} = \sqrt{m_{S,A}^2 + \mb{k}^2}.
\ee

\subsection{Entropy from $2 \to 2$ scattering}

\label{simplesection}

To begin, we study the simplest possible process: scattering with the system and apparatus both prepared in definite momentum eigenstates at early times, 
\be
\ket{\psi} = \ket{\mb{p}^+}_A \ket{\mb{q}^+}_S.
\ee
This is precisely what we studied in section \ref{entropysection} and, as we did there, we will work with box-normalized states until the end of the calculation.

After the scattering event, the von Neumann entropy of the apparatus is given directly by our formula \eqref{entropyboxformula-PT}, viz.
\be
S_A = - \sum_{\ul{\mb{p}}} F(\ul{\mb{p}}) \ln F(\ul{\mb{p}}).
\ee
Again the sum runs over all outgoing apparatus momenta $\ul{\mb{p}}$, and the coefficients $F(\ul{\mb{p}})$ are defined in \eqref{entropyboxcoeffs}. Because scattering in this theory is isotropic, it is straightforward to compute the apparatus density matrix eigenvalues explicitly. Move to the center-of-momentum frame $\mb{p} = -\mb{q}$. Then
\begin{align}
\begin{split}
F(\ul{\mb{p}}) & = 2\pi T \sum_{\ul{\mb{q}}} \left| M_{\ul{\mb{p}} \ul{\mb{q}} \mb{p} \mb{q}} \right|^2 \delta_{\ul{\mb{p}} + \ul{\mb{q}}, \mb{p} + \mb{q}} \delta(E^A_{\ul{\mb{p}}}+E^S_{\ul{\mb{q}}}-E^A_{\mb{p}}-E^S_{\mb{q}}) \\
& = 2\pi T \left| M(p_{cm}) \right|^2 \delta(f(|\ul{\mb{p}}|)), 
\end{split}
\end{align}
where we used isotropy of the interaction to write this as
\be
\label{isotropic}
M(p_{cm}) = M_{\ul{\mb{p}},-\ul{\mb{p}};\mb{p},-\mb{p}}, \ \ p_{cm} = |\mb{p}| = |\ul{\mb{p}}|, \ \ \ f(|\ul{\mb{p}}|) = E^A_{|\ul{\mb{p}}|}+E^S_{|\ul{\mb{p}}|} - E^A_{|\mb{p}|}-E^S_{|\mb{p}|}.
\ee
The entropy of $A$ at late times is thus given by
\be
\label{simpleentropybox}
S_A = - 2\pi T \sum_{\ul{\mb{p}}} \left| M(p_{cm}) \right|^2 \delta(f(|\ul{\mb{p}}|)) \ln \left[ 2\pi T \left| M(p_{cm}) \right|^2 \delta(f(|\ul{\mb{p}}|)) \right].
\ee
At this stage, we can take the continuum limit. We replace the sum $\sum_{\ul{\mb{p}}} \to V/(2\pi)^3 \int d^3\ul{\mb{p}}$, and do the integral in spherical coordinates. The delta-function outside the log enforces energy conservation, and so the delta inside the log is replaced by $\delta_T(0) = T/2\pi$. We also have to insert the appropriate factors of $\tilde{N} = (2\pi)^3/V$ to convert from the box-normalized amplitude to the continuum-normalized one, see eq. \eqref{Sdecompbox}. Finally, we obtain
\be
S_A = -2 (2\pi)^5 \frac{T}{V} p_{cm}^2 (E^A + E^S) \left| M(p_{cm}) \right|^2 \ln \left[ (2\pi)^6 \frac{T^2}{V^2} \left| M(p_{cm}) \right|^2 \right],
\ee
where the energies are understood to be evaluated at $p_{cm}$.
This holds at any order of perturbation theory. If we wanted to work to lowest order in perturbation theory, we can use our matrix element \eqref{22elements} given above, in which case we have explicitly\cite{Park:2014hya}
\be
S_A = - \frac{T}{V} \frac{\lambda^2 }{16 \pi}\frac{p_{cm} (E^A + E^S)}{(E^A E^S)^2} \ln \left[ \frac{T^2}{V^2} \frac{\lambda^2}{16 (E^A E^S)^2} \right].
\ee

This formula bears some remarking. For one thing, recall that the total cross-section for this theory at this order of perturbation theory is given by $\sigma = \lambda^2/16\pi E^A E^S$ in the center-of-momentum frame. So we have that the entropy is proportional to this quantity, integrated over time and against the flux of incoming particles.\footnote{In this frame, the flux is $\Phi = u/V$ with the relative velocity $u = p_{cm} (E^A + E^S)/E^A E^S$.} We always have a large spatial volume $V$ in mind, so $S_A \geq 0$. The argument of the logarithm likewise cannot be too small: if $T \lambda/16 V E^A E^S \leq 1$ then the entropy will be negative. This is essentially the statement that the Compton wavelengths of the particles need to be within the spacetime box. As we take the spatial volume $V \to \infty$ with $T$ fixed, $S_A$ goes to zero from above; this follows from the fact that the probability of the waves to interact at all goes to zero.  Finally, one might worry about $V$ fixed and $T \to \infty$, in which case the entropy goes to $-\infty$, but this corresponds to an infinite number of repeated interactions, which would also violate the basic assumption of the $S$-matrix setup that we are describing an isolated event.

\subsection{Verifying spatial superpositions}

\label{superpositionsection}

Let's consider now a rather different problem. Suppose we prepare the system and apparatus in a separable state, but the system state may or may not be pure. We would like to know how this system information would show up in the outgoing apparatus state. 

For definiteness, we consider the following problem: suppose that some black box machine in our lab prepares the system as either a classical ensemble or coherent superposition of two system states, each localized to a different point in real space. The question is: how do we verify the coherence of the superposition from a scattering experiment?

We will see that it is sufficient to look at the position-space wavefunction of the outgoing apparatus at order $\lambda$. The signature of the system superposition is interference fringes in the apparatus state. They show up at order $\lambda$ because the position-space projector $\ket{\mb{x}} \bra{\mb{x}}$ is sensitive to off-diagonal momentum-space apparatus density matrix elements, which are generated at first order in the perturbation, as we now demonstrate explicitly.

\begin{figure}
\centering
\includegraphics{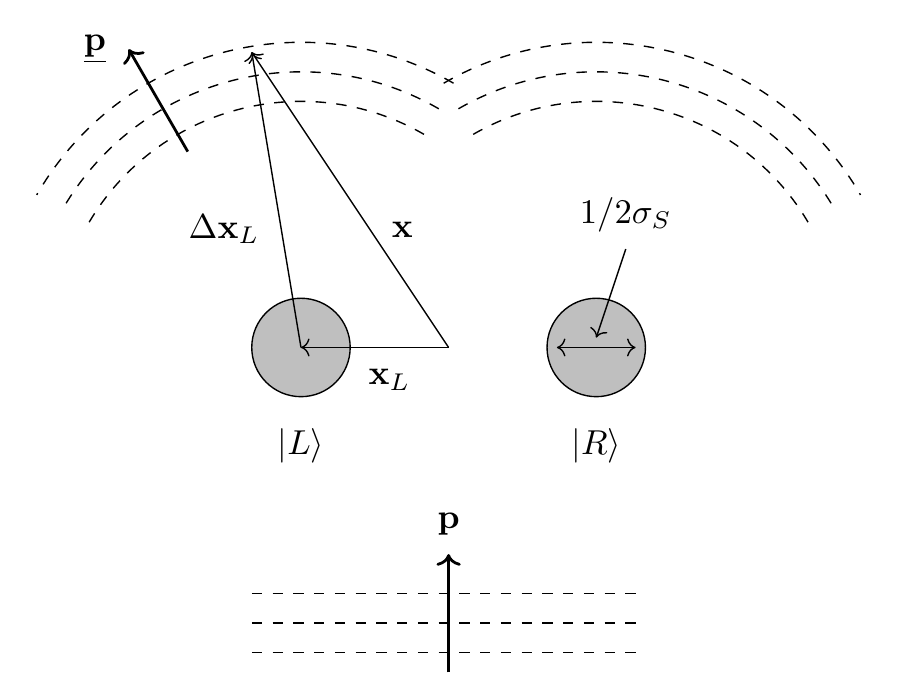}
\caption{Verifying spatial superpositions of the system states $\ket{L}$, $\ket{R}$.}
\label{LRpic}
\end{figure}

We begin by defining a pair of states $\ket{L}, \ket{R}$ that describe the apparatus prepared in an incoming state of momentum $\mb{p}$ and the system centered at different positions $\mb{x}_{L,R}$ in real space.\footnote{In this section we will use continuum-normalized states, regulating squares of Dirac deltas as
\be \left[ \delta^3(\mb{p}-\mb{p}') \right]^2 = \frac{V}{(2\pi)^3} \delta^3(\mb{p}-\mb{p}'), \ \ \ \left[ \delta(E-E') \right]^2 = \frac{T}{2\pi}\delta(E-E'). \ee } Define the usual Gaussian wavefunction
\be
g(\mb{q}) = N_S \exp \left\{ -\mb{q}^2/4\sigma_S \right\}, \ \ N_S = \frac{1}{(2 \pi \sigma_{S})^{3/4}}
\ee
and take the system to be initialized at rest in a lab frame, so we define the state as follows: let $i \in \left\{ L,R \right\}$ and put
\begin{align}
\begin{split}
\label{k-packets}
\ket{i} & = N_A \ket{\mb{p}^+}_A \int d^3\mb{q} \ f_{i}(\mb{q}) \ket{\mb{q}^+}_S, \\
f_{i}(\mb{q}) & =  g(\mb{q}) \exp \left\{ i \mb{q} \cdot \mb{x}_i \right\}, \ \ \ N_{A} = \sqrt{ \frac{(2\pi)^{3}}{V} }.\end{split}
\end{align}
See figure \ref{LRpic}. These states are not orthogonal; their overlap is 
\be
\epsilon = \braket{L | R} = \exp \left\{ -\sigma_S \left| \Delta \mb{x}_0 \right|^2/2 \right\}, \ \ \ \Delta \mb{x}_0 = \mb{x}_L - \mb{x}_R.
\ee
We have in mind that the system states are localized in real space, so that the momentum spread $\sigma_S$ is large. The two states are well-separated if $\epsilon \ll 1$; we assume this below for mathematical ease, but the results do not depend qualitatively on this condition.\footnote{When working with the following formulas, the non-orthogonality of $\ket{L},\ket{R}$ should be kept in mind; in particular traces should be done with momentum eigenstates. A useful relation is $\tr \ket{i}\bra{j} = \braket{i|j} =\epsilon$ for $i \neq j$ and $1$ for $i = j$.} We are assume that the scattering is done in a sufficiently short time so that we can ignore the spreading of these wavepackets.

Now consider an arbitrary density matrix in the space spanned by the $\ket{L},\ket{R}$ states: 
\be
\rho = \Gamma^{ij} \ket{i} \bra{j}, \ \ i,j \in \left\{ L,R \right\}.
\ee
For example, we can form a convex family of density matrices, with coefficients
\be
\label{gammacoeffs}
\Gamma^{ij}(\alpha) = \frac{1}{2(1+\epsilon)} \begin{pmatrix} 1 + \epsilon - \alpha \epsilon & \alpha \\ \alpha & 1 + \epsilon - \alpha \epsilon \end{pmatrix}, \ \ 0 \leq \alpha \leq 1.
\ee
These linearly interpolate between the classical ensemble proportional to $\ket{L} \bra{L} + \ket{R} \bra{R}$ at $\alpha = 0$ and the perfect coherent superposition proportional to $(\ket{L} + \ket{R})(\bra{L} + \bra{R})$ at $\alpha = 1$. These all have unit trace, while the purity $\tr \rho^2(\alpha) = [1 + (\alpha + \epsilon - \alpha \epsilon)^2]/2$ vanishes when $\epsilon = \alpha = 0$ and goes up to unity if either $\epsilon = 1$ or $\alpha = 1$. We will refer to $\alpha$ as the coherence parameter. Note in particular that the off-diagonal element $\Gamma^{LR}$ is linear in $\alpha$. The reduced density matrix for the apparatus expressed with out-states is
\be
\label{rhoaLRprob}
\rho_A^- = N_A^2 \sum_{ij} \int d^3\ul{\mb{p}} d^3\ul{\mb{p}}' d^3\ul{\mb{q}} d^3\mb{q} d^3\mb{q}' \Gamma^{ij} f_i(\mb{q}) f^*_j(\mb{q}') S_{\ul{\mb{p}} \ul{\mb{q}} \mb{p} \mb{q}} S^*_{\ul{\mb{p}}' \ul{\mb{q}} \mb{p} \mb{q}'} \ket{\ul{\mb{p}}^-} \bra{\ul{\mb{p}}^{'-}}.
\ee

Let's study some outgoing apparatus observables. Consider first the outgoing momentum distribution $P(\ul{\mb{p}})$ of the apparatus, so that we take $\Oi_A = \ket{\ul{\mb{p}}} \bra{\ul{\mb{p}}}$ and use \eqref{redobs3}; the expectation value can be read off from the diagonal elements of \eqref{rhoaLRprob}. We can work these out a bit more explicitly. The identity-squared term from decomposing the $S$-matrix with \eqref{Sdecomp} contributes to $P(\ul{\mb{p}})$ as $P_{0}(\ul{\mb{p}}) = \delta^3(\ul{\mb{p}}-\mb{p})$. The interaction terms give
\begin{align}
\begin{split}
\label{LRpout}
P_{int}(\ul{\mb{p}}) & = (2\pi)^3 \frac{T}{V} \int d^3\mb{q} \left| g(\mb{q}) \right|^2 \left( 1 + \alpha \cos 2 \mb{q} \cdot \Delta \mb{x}_0 \right) \Big\{ -2 \text{Im} M_{pqpq} \delta^3(\ul{\mb{p}}-\mb{p}) \\
& + \left| M_{\ul{\mb{p}}, \mb{q}-\mb{k}; \mb{p},\mb{q}} \right|^2 \delta \left(E^A_{\ul{\mb{p}}} + E^S_{\mb{q}-\mb{k}} - E^A_{\mb{p}} - E^S_{\mb{q}} \right) \Big\}
\end{split}
\end{align}
where here $\mb{k} = \ul{\mb{p}} - \mb{p}$ is the momentum transfer, and we took $\epsilon \ll 1$ to write the result in a simple way. We see that the overall probability is proportional to $T/V$, as expected. Both terms receive a contribution from the coherence $\alpha$ of the initial superposition. In our specific theory \eqref{22elements}, both of these contributions are of order $\lambda^2$, with the forward-scattering term in \eqref{LRpout} coming in only at one-loop order. So to measure $\alpha$ by doing such an observation, we would have to be sensitive at order $\lambda^2$.

\begin{figure}
\begin{center}$
\begin{array}{cc}
\includegraphics{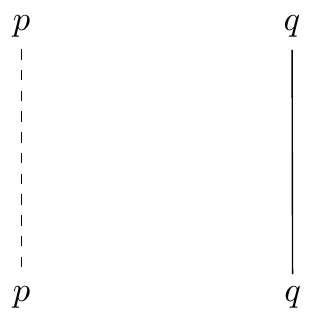} \ \ \ \
&
\includegraphics{pic-treediag.pdf}
\end{array}$
\end{center}
\caption{Diagrams contributing to the lowest-order position-space distribution of the apparatus.}
\label{positiondiagrams}
\end{figure}

However, it is possible to see signatures of the coherence $\alpha$ at \emph{first} order in $\lambda$ if we instead look at position-space observables. Consider the position-space probability distribution for the apparatus at late times after the scattering, $P(\mb{x})$. This can be obtained by again applying \eqref{redobs3} but now using the observable $\Oi_A = \ket{\mb{x}} \bra{\mb{x}}$, the single-particle position projector. The delta-squared terms from the $S$-matrix result in $P_0(\mb{x},t) = V^{-1}$ by direct computation. Next we need both the cross terms $M \rho - \rho M^{\dagger}$ and the amplitude-square $M \rho M^{\dagger}$ term; the latter will start at $\Oi(\lambda^2)$, so let us consider the former. A straightforward calculation using hermiticity of $\Gamma^{ij}$ gives
\begin{align}
\label{P1first}
\begin{split}
P_{1}(\mb{x}) & = \frac{4\pi}{V} \int d^3\mb{q} d^3\ul{\mb{q}} \delta(E^A_{\mb{p} + \mb{q} - \ul{\mb{q}}} + E^S_{\ul{\mb{q}}} - E^A_{\mb{p}} - E^S_{\mb{q}}) g(\mb{q}) g^*(\ul{\mb{q}})   \\
& \times \text{Im} \left[ M_{\mb{p}+\mb{q}-\ul{\mb{q}}, \ul{\mb{q}} ; \mb{p} , \mb{q}} \sum_{ij} \Gamma^{ij} \exp \left\{-i \phi_{ij}(\mb{q},\ul{\mb{q}}) \right\} \right],
\end{split}
\end{align}
where the subscript $1$ means we are thinking of this in first-order perturbation theory, and the phases are
\be
\phi_{ij}(\mb{q},\ul{\mb{q}}) = -E^S_{\mb{q}} t + (\mb{x} - \mb{x}_i) \cdot \mb{q} + E^S_{\ul{\mb{q}}} t - (\mb{x} - \mb{x}_j) \cdot \ul{\mb{q}}.
\ee
Consider measuring the location of the outgoing $A$ particle when $t$ and $|\mb{x}-\mb{x}_i|$ are of the same order and large. Then the integral may be approximated by its stationary phase value, which here is given when
\be
\mb{q} = \mb{q}_i = m_S \gamma_i v_i  \widehat{\Delta \mb{x}_i}, \ \ \ul{\mb{q}} = \mb{q}_j = m_S \gamma_j v_j \widehat{\Delta \mb{x}_j},
\ee
where
\be
\Delta \mb{x}_i = \mb{x} - \mb{x}_i, \ \ \widehat{\Delta \mb{x}_i} = \frac{\Delta \mb{x}_i}{| \Delta \mb{x}_i|}, \ \ v_i = \frac{|\Delta \mb{x}_i|}{t}, \ \ \gamma_i = \frac{1}{\sqrt{1-v_i^2}}.
\ee
Note that $v_i = v_i(\mb{x},t)$ and likewise $\gamma_i = \gamma_i(\mb{x},t)$ depend on the point of observation $\mb{x}$ and the time $t$; we suppress this dependence in the formulas that follow. At these values for the momenta, we have that
\be
E^S_{q_i} = m_S \gamma_i, \ \ \ \phi_{ij} = -m_S t \left[ \gamma_i^{-1} - \gamma_j^{-1} \right] = -\phi_{ji}.
\ee
In particular, we see that the $LL$ and $RR$ terms have zero phase, and thus give real contributions in \eqref{P1first} since our amplitude \eqref{22elements} is real at lowest order, so they do not contribute to the outgoing position distribution. The interference terms $LR$ and $RL$ do contribute, however, and we get
\be
\label{P1final}
P_1(\mb{x},t) = A(\mb{x},t) \sin \left( \phi_{LR}(\mb{x},t) \right)
\ee
where at this point we have finally used the reality of our amplitude \eqref{22elements}. The position-space amplitude is
\be
A = \alpha \frac{2 (2\pi)^4}{V} \gamma_L^{5/2} \gamma_R^{5/2} m_S^3 t^{-3} g(\mb{q}_L) g(\mb{q}_R) \left[ \delta_{LR} M_{LR} - \delta_{RL} M_{RL} \right] 
\ee
where we defined for brevity
\be
M_{ij} = M_{\mb{p}_0 + \mb{q}_i - \mb{q}_j, \mb{q}_j ; \mb{p}_0, \mb{q}_i}, \ \ \ \delta_{ij} =  \delta(E^A_{\mb{p} + \mb{q}_i - \mb{q}_j} + E^S_{\mb{q}_i} - E^A_{\mb{p}} - E^S_{\mb{q}_j}).
\ee
The delta-functions localize the distribution to the stationary-phase wavefronts, and are an artifact of the way we did the integrals. In reality, they should be smoothed out.

The key physics is in the sine term in \eqref{P1final}, and the fact that $A$ is linear in both the coupling $\lambda$ and coherence parameter $\alpha$. The amplitude $A$ is a rather complicated function of $\mb{x},t$, but the point is clear enough: if we arrange an array of particle detectors in a sphere around the origin, it will pick up the interference pattern given by the sine term in \eqref{P1final}. The heights of the interference fringes, in turn, are set by the coherence $\alpha$: in particular, if the system is initialized in a classical ensemble, $\alpha = 0$ and there are no fringes. 

Physically, these are interferences between the process where no scattering occurs and the process where the apparatus scatters off one or the other system locations, see figure \ref{positiondiagrams}. Mathematically, this is in the $M \rho 1 - 1 \rho M^{\dagger}$ terms in the action of the $S$-matrix on the density matrix. This is why the interference appears at order $\lambda$ and not $\lambda^2$. This should be contrasted with momentum-space observables, which are only sensitive to the interference at $\lambda^2$: the position-space observable is sensitive to \emph{off-diagonal} momentum-space density matrix elements, which are generated at lowest order in perturbation theory.

\section{Conclusions}
We have studied some prototypical examples of an apparatus particle scattering off a collection of system particles, applying the language of quantum measurement theory to a field-theoretic problem. Our general density matrix formalism allows for the computation of arbitrary apparatus observables at early and late times, and we showed how to compute the apparatus-system entanglement entropy generated during scattering.

Our scenario contrasts standard formulations of measurement theory in some significant ways. For one thing, our system and apparatus are relativistic and have continuous spectra. For another, we do not imagine that we can precisely engineer some interaction Hamiltonian; here we are just stuck with whatever our effective field theory happens to give us. Nonetheless we have found that it is straightforward to use standard measurement-theory techniques.

A potential application is detection of system properties at lower orders of perturbation theory than usually considered in scattering. For example, one often hears that $\lambda \phi^4$ scattering is only sensitive to $\lambda^2$ as opposed to $\lambda$, because the cross-section scales like $\lambda^2$. On the contrary, one can clearly do an interference measurement as described above to measure the coupling at order $\lambda$.

More theoretically, these kinds of calculations may help shed some light on certain aspects of black hole physics. In particular, a recent proposal is that the black hole information is radiated out to null infinity by soft bosonic modes.\cite{Hawking:2016msc} This information should thus be quantified by precisely the kind of von Neumann entropy we have considered here. Implications of the soft boson theorems for the entropy calculations presented above will appear in a future article.

\section*{Acknowledgements}
We thank Charles Rabideau, Dennis R\"{a}tzel, Jess Riedel, Philip Stamp, Bill Unruh, and Jordan Wilson for discussions. We used the TikZ-Feynman package \cite{Ellis:2016jkw} to make our Feynman diagrams. We acknowledge that our work at UBC was performed on the unceded territory of the Musqueam people. All three of us are grateful for support from NSERC, and DC from the Templeton Foundation award 36838 and the Pacific Institute of Theoretical Physics.

\appendix
\section{Optical theorem}
\label{opticaltheoremapp}

Here we repeat Weinberg's proof of the optical theorem, for completeness, because the same techniques appear repeatedly in the above. In particular, we explain how unitarity of the density matrices used in scattering is directly related to the optical theorem.

Our scattering states are supposed to be continuum-normalized
\be
\braket{ \alpha^{'\pm} | \alpha^{\pm}} = \delta(\alpha-\alpha'),
\ee
where the right hand side as usual means a product of Dirac deltas on the spatial momenta. Now, this equation needs to be consistent with the unitarity of the $S$-matrix, i.e. we should have
\be
\delta(\alpha-\alpha') = \braket{ \alpha^{'+} | \alpha^{+}}  = \int d\beta d\beta' S_{\beta\alpha} S^*_{\beta' \alpha'} \braket{ \beta^{'-} | \beta^{-}} = \int d\beta S_{\beta \alpha} S^*_{\beta \alpha'}.
\ee
Writing the usual decomposition of $S$ as in \eqref{Sdecomp} and doing some of the integrals, we see that we need
\begin{align}
\begin{split}
& 2\pi i \left[ M_{\alpha' \alpha} \delta^4(p_{\beta} - p_{\alpha}) - M^*_{\alpha \alpha'} \delta^4(p_{\beta} - p_{\alpha'}) \right] \\
& = (2\pi)^2 \int d\beta M_{\beta \alpha} M^*_{\beta \alpha'} \delta^4(p_{\beta} - p_{\alpha}) \delta^4(p_{\beta} - p_{\alpha'}).
\end{split}
\end{align}
Specialize to the case $\alpha = \alpha'$. We obtain the optical theorem
\be
\label{opticaltheorem}
\text{Im} M_{\alpha \alpha} = -\pi \int d\beta \left| M_{\beta \alpha} \right|^2 \delta^4(p_{\beta} - p_{\alpha}).
\ee

Consider scattering an initial state $\ket{\alpha^+}^{box}$, now in a finite spacetime box as described in the main text. Then our density matrix should have unit trace. Writing this after applying the $S$-matrix and doing the trace using out-states $\ket{\beta^-}^{box}$, we have
\be
1 = \tr \rho = \sum_{\beta} \left| S^{box}_{\beta \alpha} \right|^2.
\ee
If we expand the $S$-matrix as in \eqref{Sdecompbox}, then the delta-squared term on the right hand side  will give exactly the $1$ on the left-hand side in our trace norm condition here. So then the remaining three terms will have to cancel amongst themselves, which is exactly the case when \eqref{opticaltheorem} holds.

Note that in perturbation theory in some weak coupling $\lambda$, the optical theorem mixes orders of $\lambda$. For our purposes above, for example to get the entanglement entropy to $\Oi(\lambda^2)$, we need to ensure that we normalize the density matrix to $\tr \rho = 1 + \Oi(\lambda^3)$. But then we need the scattering matrix elements appearing in \eqref{opticaltheorem} to cancel on the two sides of the equation up to $\Oi(\lambda^2)$. In other words, to explicitly check the normalization of the density matrix in perturbation theory at this order, we need to include the lowest-order loop diagram for forward scattering in computing the scattering amplitudes.

\bibliographystyle{utphys-dan} 
\bibliography{swpi-2}

\end{document}